\documentclass[conference]{IEEEtran}
\IEEEoverridecommandlockouts
\usepackage{cite}
\usepackage{amsmath,amssymb,amsfonts}
\usepackage{algorithmic}
\usepackage{graphicx}
\usepackage{textcomp}
\usepackage{xcolor}
\usepackage{booktabs}
\def\BibTeX{{\rm B\kern-.05em{\sc i\kern-.025em b}\kern-.08em
    T\kern-.1667em\lower.7ex\hbox{E}\kern-.125emX}}
    \usepackage[letterpaper, left=0.625in, right=0.625in, bottom=1in, top=0.75in]{geometry}

\begin{document}

\title{Temporal Derivative Soft-Sensing and Reconstructing Solar Radiation and Heat Flux from Common Environmental Sensors}
\author{Neksha DeSilva \\ Independent Researcher\\ FIA Operating Systems\\ Colombo, Sri Lanka\\ \text{n.desilva@fiaos.org}\\ \\
\text{Dated: 23rd of December, 2025}
}

\maketitle

\begin{abstract}
Modern methods of environmental monitoring are deficient in the lack of ability to take measurements of energy flows since traditional readings involve capturing parameters such as temperature, pressure, and humidity without considering their physical causes. The present research describes Differential Temporal Derivative Soft-Sensing (DTDSS), a physics-based approach which enables any ordinary low cost sensor array to infer estimates of the energy exchange in the environment by modeling its radiative heat fluxes. In particular, the proposed approach combines a novel paired sensor configuration along with a unique algorithmic solution called Inertial Noise Reduction or INR, that mathematically models the flow of energy in the environment by computing Global Horizontal Irradiance, or GHI, and convective heat flux. Experimental field testing has been conducted with the use of calibrated reference pyranometers supplied by the Department of Meteorology of Sri Lanka, yielding a correspondence between 8-bit embedded processor results and the reference of R2 > 0.9 and RMSE approx. eqv. to 45 Watts per square meter in under 2KB RAM of a microcontroller unit.
\end{abstract}


\section{Introduction}

Environmental monitoring through common sensors such as the Bosch BME280 is conventionally limited to static variables: temperature, pressure, and humidity \cite{bme280}. While these sensors provide instantaneous state data, they remain unresponsive to dynamic energy exchanges. Specifically, commercial IoT nodes rarely measure:

\begin{enumerate}
\item Solar energy reaching the ground per unit area, specifically Global Horizontal Irradiance (GHI).
\item The rate of thermal energy transfer through the environment, referred to as Heat Flux.
\end{enumerate}

These metrics are essential for practical applications. Solar radiation serves as the primary energy input for Earth's climate and is required for calculating evapotranspiration in agriculture \cite{allen1998}.

In the development of low-cost distributed sensing networks, initial experiments utilizing Thermoelectric Generator (TEG) modules to implement the Penman-Monteith equation proved unsuccessful. Experimental results indicated that the ceramic construction of standard TEGs exhibited excessive thermal inertia, retaining heat rather than providing the responsive data required for real-time flux estimation.

To address this, a cohesive, integrated approach was developed using the Bosch BME280 as a baseline foundation. Its accuracy in measuring three physically connected parameters, namely Relative Humidity, Temperature, and Pressure, provides the stability needed for an accumulative sensing package. Notably, the measurement of absolute barometric pressure allows for the direct derivation of air density, avoiding altitude-derived estimation errors common in standard atmosphere models.

\subsection{Limitations of Current Instrumentation}

Solar radiation measurement traditionally requires thermopile pyranometers governed by ISO 9060 standards \cite{iso9060} or semiconductor photodiodes. While photodiodes are inexpensive, they suffer from inherent long-term deployment challenges.

First, polymer lenses and optical epoxies degrade under prolonged UV exposure, altering spectral transmission properties. Second, optical sensors are highly susceptible to fouling; dust accumulation or biofilm growth causes immediate signal attenuation. Conversely, a thermodynamic approach measures total energy absorption rather than photon flux. A blackened thermal absorber continues to function effectively even with minor surface contamination, as it measures the heat generated by incident energy. This offers a distinct advantage in remote deployments where regular cleaning is impossible.

\subsection{Challenges in Edge-Based Soft-Sensing}

Current soft-sensing approaches typically rely on machine learning (ML). While recent studies have successfully employed LSTM and XGBoost for irradiance forecasting \cite{SolarML2023}, these models typically require significant memory overhead. This approach presents specific challenges for embedded devices:

\begin{enumerate}
\item As noted by Ray et al. \cite{TinyML2024}, deploying inference models on edge devices faces strict energy and memory boundaries, necessitating algorithmic efficiency over raw model complexity.
\item Neural networks trained in specific climates often fail to generalize to new environments without retraining.
\item Standard ML models often map instantaneous inputs to outputs, ignoring the crucial temporal dynamics of how a sensor heats up over time.
\end{enumerate}

Radiation changes on timescales of seconds (cloud movement), while temperature changes on timescales of minutes. This temporal mismatch introduces significant error in static mapping models.

\subsection{The Differential Thermodynamic Approach}

To resolve these limitations, Differential Temporal Derivative Soft-Sensing (DTDSS) is proposed. A fundamental thermodynamic principle states that a single sensor cannot simultaneously achieve thermal equality with air (for temperature/humidity measurement) and thermal disequilibrium with solar radiation (for radiation measurement). Sensing must therefore be partitioned into two separate thermodynamic nodes \cite{samani2000}.

This aligns with recent frameworks in ``Physics-Enhanced TinyML'' \cite{PhysicsTinyML2024}, demonstrating that integrating domain knowledge (thermodynamics) with sensing algorithms significantly outperforms purely data-driven black-box models in unpredictable environments.

Two identical sensors such as the BME280 are placed in close proximity but in contrasting enclosures:

\begin{enumerate}
\item The reference node ($S_{ref}$) is housed in a ventilated, reflective shield, maintaining thermal equilibrium with ambient air.
\item The flux node ($S_{flux}$) is housed in a sealed, absorptive black-body enclosure, acting as an energy trap.
\end{enumerate}

By analyzing real-time variation between these nodes, specifically the temporal derivative of their temperature difference, incident radiation can be mathematically reconstructed without optical components.

\subsection{System Architecture}

\begin{figure}[htbp]
\centerline{\includegraphics[width=\columnwidth]{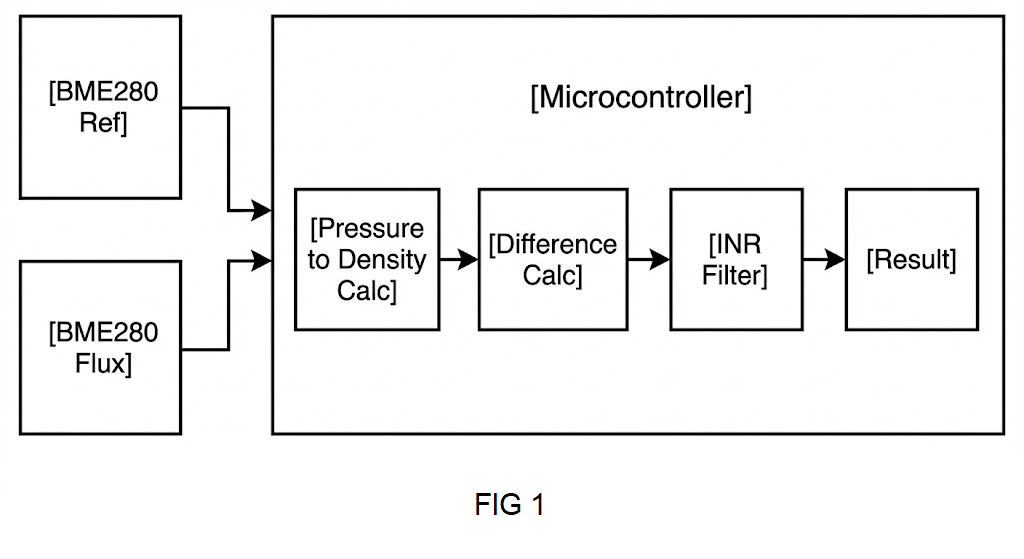}} 
\caption{Signal processing pipeline showing the sequential integration of real-time air density calculation and Inertial Noise Reduction (INR).}
\label{fig:pipeline}
\end{figure}

The pipeline sequentially executes four stages: pressure-to-density calculation, differential temperature calculation, INR filtering, and final flux reconstruction. This sequential integration is illustrated in Fig. \ref{fig:pipeline}.

The differential system comprises two processing paths.

The Reference Path operates on the vented sensor, where psychrometric relationships and empirical meteorology establish accurate baseline conditions. True ambient temperature ($T_{ref}$), relative humidity, and pressure readings are provided uncontaminated by solar heating. This path provides stability and ensures the system has accurate environmental context, preventing the so-called ``dry kiln'' error, which occurs when an enclosed sensor reports artificially low humidity due to localized self-heating, a bias that DTDSS cancels through its differential topology.

The Reactive Path operates on the black-body sensor, where the temporal derivative of temperature difference is used to detect and quantify radiative energy input, capturing transient events such as cloud breaks.

Central to the Reactive Path is Inertial Noise Reduction (INR), a signal processing technique designed for sensor thermodynamics, addressing complications arising from differentiating noisy signals.

Using the BME280's pressure sensor for real-time air density calculation, the system becomes altitude-independent, deployable from sea level to 4000+ meters without modification.

\section{The Differential Architecture}

The central challenge is distinguishing between a hot day (high ambient temperature) and a sunny day (high radiative flux). A single sensor cannot separate these signals. The differential system resolves this by processing two concurrent telemetry streams from physically distinct sensor nodes.

\begin{table}[h]
\caption{Comparison of Differential Pipeline Characteristics}
\centering
\resizebox{\columnwidth}{!}{%
\begin{tabular}{|l|l|l|}
\hline
\textbf{Feature} & \textbf{Reference Path} & \textbf{Reactive Path} \\
\hline
Time Scale & Minutes to Hours & Seconds \\
Physics Model & Equilibrium Thermodynamics & Non-Equilibrium \\
Primary Equation & Kasten-Czeplak & Newton's Law (Inverted) \\
Noise Sensitivity & Low (Heavy Filtering) & High (Requires INR) \\
Failure Mode & Slow reaction & Sensitive to sensor noise \\
\hline
\end{tabular}%
}
\end{table}

\section{Thermodynamics of Humid Air}

The sensor operates immersed in the lower atmosphere; atmospheric properties govern heat transfer between sensor and environment. Engineering models often treat air as a simple ideal gas with constant properties, or assume standard atmosphere conditions. For a flux model intended for global deployment, these simplifications introduce unpredictable outputs.

DTDSS rejects static assumptions in favor of dynamic, first-principles derivation of air properties from measured state variables: pressure ($P$), temperature ($T$), and relative humidity ($RH$).

\subsection{Air Density}

Air density ($\rho$) governs convective heat transfer \cite{incropera2007}. When the sensor is warmer than surrounding air, heat flows via convection at a rate dependent on fluid density.

At sea level, $\rho \approx 1.225$ kg/m$^3$; at 2000m, $\rho \approx 1.00$ kg/m$^3$; at 4000m, $\rho \approx 0.82$ kg/m$^3$. If sea-level density is assumed at 4000m, convective cooling is overestimated by approximately 50\%, causing proportional underestimation of solar radiation.

The BME280 provides total atmospheric pressure $P$, directly related to altitude. Actual air density is calculated without GPS or altitude lookup tables.

\subsection{Ideal Gas Law for Moist Air}

The atmosphere is a mixture of dry air and water vapor with different molecular weights and gas constants. By Dalton's Law of Partial Pressures:

\begin{equation}
P = P_d + e
\end{equation}

where $P_d$ is dry air partial pressure and $e$ is water vapor partial pressure. Moist air density:

\begin{equation}
\rho_{moist} = \frac{P - e}{R_d T} + \frac{e}{R_v T}
\end{equation}

where $R_d = 287.058$ J·kg$^{-1}$·K$^{-1}$ (dry air) and $R_v = 461.495$ J·kg$^{-1}$·K$^{-1}$ (water vapor). Defining $\epsilon = R_d/R_v \approx 0.622$:

\begin{equation}
\rho_{moist} = \frac{P}{R_d T} \left( 1 - \frac{e}{P} (1 - \epsilon) \right)
\end{equation}

Humid air is less dense than dry air at constant temperature and pressure due to the lower molecular weight of water vapor ($18 \text{ g/mol}$) compared to dry air ($\approx 29 \text{ g/mol}$). For soft sensing, this means humid air is less efficient at carrying heat away from the sensor.

By calculating $\rho_{moist}$ dynamically at every timestep, the soft sensor recalibrates its thermal model for exact altitude and humidity conditions, directly rejecting the ``standard atmosphere'' constraint.

\subsection{Vapor Pressure Derivation}

Water vapor partial pressure $e$ is derived from relative humidity and saturation vapor pressure \cite{alduchov1996}:

\begin{equation}
e = e_s(T) \cdot \frac{RH}{100}
\end{equation}

Saturation vapor pressure approximation:

\begin{equation}
e_s(T) = 6.112 \cdot \exp\left( \frac{17.67 \cdot (T - 273.15)}{T - 29.65} \right)
\end{equation}

where $T$ is in Kelvin and $e_s$ in hPa. This equation is robust for -40°C to +50°C ranges.

\subsection{Specific Enthalpy of Moist Air}

Enthalpy represents total energy content, encompassing sensible heat (thermometer measurement) and latent heat (energy stored in water vapor phase). Specific enthalpy per unit mass of dry air:

\begin{equation}
h = h_a + x \cdot h_v
\end{equation}

where $x = 0.622 \frac{e}{P - e}$ is the humidity mixing ratio. Individual enthalpy terms:
\begin{align}
h_a &\approx 1.006 \, t \text{ (kJ/kg)} \\
h_v &\approx 2501 + 1.86 \, t \text{ (kJ/kg)}
\end{align}

The value 2501 kJ/kg represents latent heat of vaporization at 0°C. The total specific enthalpy:

\begin{equation}
h = 1.006 \, t + x(2501 + 1.86 \, t)
\end{equation}

The Reference Path tracks enthalpy changes to distinguish between Advective heating (warm, humid air moving in) and Radiative heating (sunlight hitting the sensor).

To illustrate the necessity of the enthalpy model, consider two states:
\begin{enumerate}
    \item In state A, temperature rises from 20\textdegree C to 25\textdegree C with constant 50\% humidity.
    \item In state B, temperature rises from 20\textdegree C to 25\textdegree C, but humidity increases from 40\% to 70\%.
\end{enumerate}
A standard scalar temperature model treats these as identical. However, state B represents a significantly higher change in total environmental energy due to the latent heat carried by the additional water vapor. To put the scale of this energy reservoir in perspective, evaporating a single gram of water requires as much energy as heating that same gram by 2501\textdegree C, were such heating possible in the liquid phase. The DTDSS framework identifies this distinction through the Reference Path, distinguishing advective weather fronts from true radiative events.

\subsection{Convective Heat Transfer Coefficient}

The convective heat transfer coefficient $h_c$ dictates heat flux rate:

\begin{equation}
q_{conv} = h_c (T_s - T_{\infty})
\end{equation}

For small objects like BME280 packages (approximately 2.5mm × 2.5mm × 0.9g), flow is typically laminar at low wind speeds. The Nusselt number relates $h_c$ to thermal conductivity:

\begin{equation}
Nu = \frac{h_c \cdot L}{k_{air}}
\end{equation}

Empirical correlations for laminar flow: $Nu \propto Re^{1/2} \cdot Pr^{1/3}$. Expanding:

\begin{equation}
h_c \propto k_{air} \left( \frac{\rho V}{\mu} \right)^{0.5}
\end{equation}

Thus $h_c$ depends on $\rho^{0.5}$, making altitude correction essential. As altitude increases, $\rho$ decreases, $h_c$ decreases, and the sensor becomes better isolated by thinner air. Without this correction, higher temperature rise from poor cooling would be interpreted as increased solar radiation, reporting erroneous ``phantom solar events'' every time the sensor is taken up a mountain.

By driving the model with calculated $\rho_{moist}$, DTDSS normalizes heat flux estimation against altitude variations.

\section{Signal Processing Architecture}

\subsection{Pipeline A: The Reference Path}

The Reference Path takes input from the ventilated sensor ($S_{ref}$) and operates within the domain of equilibrium thermodynamics.

This pipeline establishes ground truth of the air mass, providing accurate $T_{ref}$, $RH$, and $P$. The Reference Node provides true ambient temperature $T_{\infty} = T_{ref}$ directly without estimation.

Cloud Proxy Model (Kasten-Czeplak Adaptation): High surface relative humidity correlates with cloud formation and atmospheric opacity \cite{kasten1976}. Lacking a sky camera, a proxy is derived from relative humidity. It is important to note that $G_{cs}$ (clear-sky radiation) is not a static constant; it is calculable from latitude, longitude, time of day, and day of year, allowing the system to know its theoretical maximum at any moment. Observational data suggests a power-law relationship:

\begin{equation}
N_{proxy} = \left( \frac{RH}{100} \right)^k
\end{equation}

where $k$ is a tuning parameter (typically 1.5–2.0). The Reference Path radiation estimate then becomes:

\begin{equation}
\resizebox{0.85\columnwidth}{!}{
$G_{baseline} = G_{cs} \cdot \left( 1 - 0.75 \cdot \left( \frac{RH}{100} \right)^{3.4k} \right)$
}
\end{equation}

\subsection{Pipeline B: The Reactive Path}

The Reactive Path takes input from the black-body sensor ($S_{flux}$) and operates within the domain of non-equilibrium dynamics.

This sensor is permitted to overheat. Deviation of $T_{flux}$ from $T_{ref}$ is monitored. The temporal derivative of temperature difference identifies rapid heating events.

While standard datasheets suggest minimizing self-heating (0.5--2.0\textdegree C), the DTDSS framework embraces this thermal coupling for the Flux Node. By quantifying this ``defect,'' a common error source transforms into a diagnostic signal. When the Flux sensor heats faster than the Reference, the excess energy is definitively attributed to external radiative forcing.

\subsection{Pipeline Interaction}

The two pipelines interact:
\begin{enumerate}
\item The Reference Path provides baseline $T_{ref}$ for the Reactive Path, specifically as a direct measurement rather than an estimate.
\item The Reference Path provides sanity bounds. If the Reactive Path calculates 1500 W/m$^2$ but the clear-sky maximum is 900 W/m$^2$, an anomaly is flagged.
\item Common-mode noise rejection is achieved because both sensors experience the same wind, pressure changes, and air mass movements. Computing the differential ($T_{flux} - T_{ref}$) cancels common disturbances, isolating the solar signal.
\end{enumerate}

The interaction is bidirectional: if the Reactive Path detects a sharp flux increase while the Reference Path (based on humidity) predicts cloud cover, the system can dynamically revise the cloud proxy parameter $k$ to better match local atmospheric opacity. This feedback loop enables real-time self-correction without manual intervention.

\section{Mathematical Derivation of Heat Flux}

Temperature differential is converted to heat flux by treating the Flux sensor package as a Lumped Capacitance System, valid when Biot number ($Bi$) is less than 0.1.

\subsection{Differential Energy Balance}

Conservation of energy applied to the Flux Node:

\begin{equation}
(\alpha G A_s + P_{elec}) - h_c A_s (T_{flux} - T_{ref}) = m C_p \frac{dT_{flux}}{dt}
\end{equation}

where $\alpha$ is solar absorptivity, $G$ is solar radiation, $A_s$ is surface area, $m C_p$ is heat capacity, and $h_c$ is heat transfer coefficient.

\subsection{Solving for Solar Radiation}

Dividing by $h_c A_s$ and defining $\tau = \frac{m C_p}{h_c A_s}$ (thermal time constant) and $T_{rise} = \frac{P_{elec}}{h_c A_s}$ (self-heating offset), this first defines the Sol-Air Excess ($T_{sol}$) as the isolated thermal signal representing external forcing:

\begin{equation}
T_{sol} = (T_{flux} - T_{ref}) + \tau \frac{dT_{flux}}{dt} - T_{rise}
\end{equation}

This term allows for the calculation of Solar Radiation ($G$):

\begin{equation}
G = \frac{h_c}{\alpha} T_{sol}
\end{equation}

\subsection{Interpretation and Newton's Law Inversion}

The system behavior splits into two distinct modes:

\begin{enumerate}
    \item During steady state ($\frac{dT}{dt} = 0$), measurement relies solely on the temperature difference ($T_{flux} - T_{ref}$), indicating equilibrium.
    \item During transient state ($\frac{dT}{dt} > 0$), when the sun emerges, the term $\tau \frac{dT}{dt}$ represents the Inertial Correction, accounting for energy currently stored in the sensor's mass that has not yet appeared as a full temperature rise.
\end{enumerate}

The inclusion of the derivative term essentially inverts Newton's Law of Cooling. While Newton's Law states that the rate of cooling is proportional to the temperature difference ($\frac{dT}{dt} \propto \Delta T$), this framework rearranges the equation to solve for the magnitude of the external forcing required to cause the observed rate of change.

\section{Inertial Noise Reduction (INR)}

The derivation relies on instantaneous derivative $\frac{dT_{flux}}{dt}$. In digital sampling, differentiation is problematic \cite{savitzky1964}.

\subsection{The Noise Problem}

BME280 temperature resolution is approximately 0.01°C with quantization noise at this resolution. For a sensor with a resolution of $0.01^{\circ}$C and a sampling interval of 1s, a single bit-flip produces a derivative of $0.01^{\circ}$C/s. If the thermal time constant $\tau$ is 30s, this spike is amplified to an effective temperature error of $0.3^{\circ}$C, which may be interpreted as a radiation flux of tens of W/m$^2$. This creates a sampling paradox for constrained hardware: sampling faster amplifies derivative noise ($\text{Noise}_{derivative} \propto 1/\Delta t$), while sampling too slowly risks missing highly transient events, such as rapid cloud gap transitions.

\subsection{INR Filter Concept}

The sensor itself acts as a physical filter, with thermal mass smoothing rapid radiation changes. INR models physical inertia digitally.

The first stage employs adaptive smoothing through an Infinite Impulse Response (IIR) filter structure based on an Exponential Moving Average (EMA). The smoothing factor $\alpha$ is made adaptive based on the jerk of the signal, following this logical sequence:

\begin{enumerate}
    \item Compute deviation: $\Delta[n] = |T_{raw}[n] - T_{filt}[n-1]|$
    \item Adjust alpha: $\alpha[n] = \text{clamp}(k \cdot \Delta[n], \alpha_{min}, \alpha_{max})$
    \item Apply the adaptive EMA to generate $T_{filt}$.
    \item Calculate the slope using a central difference to obtain the derivative.
    \item Generate $T_{proj}$ via inertial projection to find the zero-mass state.
\end{enumerate}

This allows the system to increase inertia during steady states to lock values, and decrease inertia during transient events to track edges accurately.

The second stage performs inertial projection, where the projected temperature compensates for thermal lag:

\begin{equation}
T_{proj}[n] = T_{filt}[n] + \tau \cdot \left( \frac{dT_{filt}}{dt} \right)[n]
\end{equation}

The projected temperature represents a ``virtual'' zero-thermal-mass sensor. When radiation spikes, $T_{proj}$ immediately jumps toward the new equilibrium, while the physical $T_{filt}$ climbs slowly; the difference between them represents the latent energy currently being stored in the sensor's physical mass.

For derivative computation, Savitzky-Golay filtering or central difference is used.
\begin{equation}
\left(\frac{dT}{dt}\right)[n] = \frac{T_{filt}[n+1] - T_{filt}[n-1]}{2 \Delta t}
\end{equation}

While central difference is computationally optimal for 8-bit systems, the DTDSS framework also supports Savitzky-Golay filtering for derivative computation. This technique fits a low-degree polynomial to a sliding window of data points to calculate the derivative analytically, further suppressing high-frequency quantization artifacts without introducing phase lag \cite{savitzky1964}.

\subsection{INR vs. Kalman Filtering}

Kalman filters have disadvantages for this application: Computational cost of matrix operations on 8-bit MCUs and difficulty in tuning process noise without ground truth. INR requires scalar arithmetic only, is O(1) in time/space, and is specifically optimized for sensor package physics.

\section{Wind and Boundary Conditions}
\begin{figure}[htbp]
\centerline{\includegraphics[width=\columnwidth]{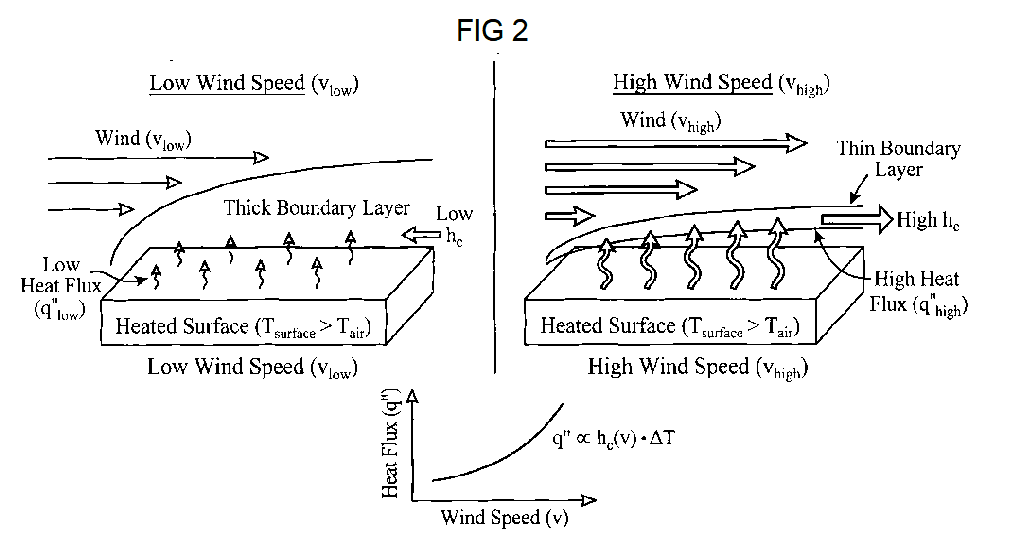}} 
\caption{The Correlation Between Wind Performance and Heat Flux}
\label{fig:performance}
\end{figure}

Wind speed directly controls convective heat transfer coefficient ($h_c$). In forced convection, $h_c$ can be 5–10 times higher than in natural convection. As visualized in Fig. \ref{fig:performance}, the correlation between wind speed and heat flux variance confirms the necessity of the compensation algorithm.

\subsection{Limitations of Differential Cancellation}

A common misconception is that differential sensing ($T_{flux} - T_{ref}$) automatically cancels wind noise. Thermodynamically, this is inaccurate. According to Newton's Law of Cooling, the rate of energy loss is proportional to temperature difference $(T_s - T_{air})$. Since the Flux sensor is significantly hotter than the Reference sensor, a sudden gust of wind removes more absolute energy (Joules) from the Flux sensor than the Reference sensor.

The system addresses this not just through subtraction, but through frequency discrimination. Wind gusts typically manifest as high-frequency noise ($>1$ Hz), whereas solar radiative forcing drives lower-frequency thermal ramps ($<0.1$ Hz). The INR filter is tuned to reject the high-frequency convective ``chatter'' caused by gusts while passing the lower-frequency solar derivative signal.

\subsection{Local Convective Correlation}

\begin{figure}[htbp]
\centerline{\includegraphics[width=\columnwidth]{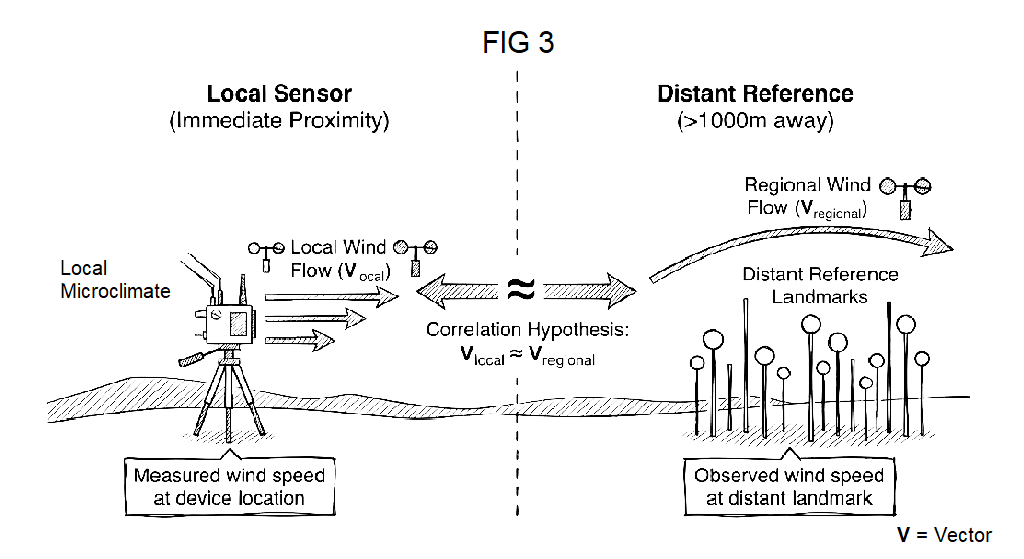}} 
\caption{Proximity Correlation Hypothesis: Local sensor-level wind speed correlates with regional wind flow patterns.}
\label{fig:wind}
\end{figure}

Practical testing observed that local wind speed at the sensor location frequently correlates with broader air mass movements in open environments. This is supported by observations where local sensor-level wind variance showed correlation with visible regional wind markers (e.g., foliage movement on landmarks) located over 1,000m away (see Fig. \ref{fig:wind}). This implies that in open environments, local wind speed correlates with the general area, enabling the use of cloud-based weather data to estimate local convective cooling where on-device anemometers are absent.

\subsection{Dual-Compartment Enclosure}

\begin{figure}[htbp]
\centerline{\includegraphics[width=\columnwidth]{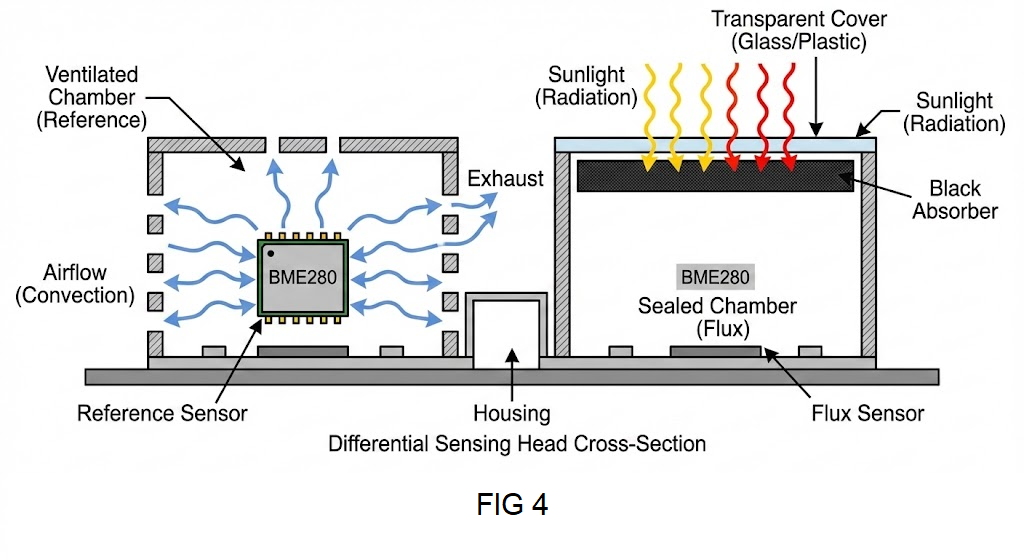}} 
\caption{Differential Sensing Head Cross-Section. The Flux node is isolated under a transparent dome, while the Reference node is ventilated.}
\label{fig:cross_section}
\end{figure}

Physical implementation uses a dual-compartment design. As illustrated in Fig. \ref{fig:cross_section}, the Flux sensor is placed beneath a transparent cover (glass or polyethene) and backed by a matte black carbon-coated aluminum absorber. The use of aluminum ensures high thermal conductivity, validating the Lumped Capacitance assumption by minimizing internal temperature gradients ($Bi < 0.1$). Meanwhile, the Reference sensor remains in a ventilated chamber to track ambient air mass movement independently. Crucially, the thermal mass of the two sensors should be kept identical. This symmetry ensures that ``common-mode noise,'' such as a sudden gust of cold wind, affects both nodes at the same rate. This allows the differential calculation ($T_{flux} - T_{ref}$) to subtract out wind noise more effectively, isolating the true solar signal.

\subsection{Limitations of Correlation in Different Environments}
The correlation hypothesis holds primarily in open outdoor environments where large-scale pressure gradients drive uniform wind flow. Indoors, microclimates (e.g., HVAC systems or greenhouses) create air movements with zero correlation to regional weather, necessitating the disabling of the wind-correction model.

\section{Calibration and Deployment}

\subsection{Essential Parameters}

\begin{table}[h]
\centering
\begin{tabular}{|l|c|c|l|}
\hline
\textbf{Parameter} & \textbf{Symbol} & \textbf{Value} & \textbf{Method} \\
\hline
Thermal time constant & $\tau$ & 10–60 s & Auto-calibration \\
Solar absorptivity & $\alpha$ & 0.85–0.95 & Literature \\
Surface area & $A_s$ & ~50 mm$^2$ & Geometry \\
Heat capacity & $m C_p$ & ~1 J/K & Calculation \\
Self-heating rise & $T_{rise}$ & 0.5–2.0°C & Dark measurement \\
\hline
\end{tabular}
\end{table}

\subsection{Auto-Calibration and Self-Heating}

Thermal time constant $\tau$ is determined via step response test by identifying when cooling starts and measuring the time to reach 63\% of the temperature drop. The value of $\tau$ can be updated periodically (e.g., monthly) to account for dust accumulation on the housing. Virtual sensing architectures often rely on cross-correlating auxiliary parameters \cite{VirtualSensing2025} to maintain accuracy without manual intervention, a principle adapted for the DTDSS auto-calibration routine. 

Self-heating characterization ($T_{rise}$) is determined by allowing the dual-sensor system to reach equilibrium in a dark, zero-radiation environment, creating a lookup table to account for varying power dissipation at different sampling rates.

\subsection{Failure Modes and Sanity Checks}
To ensure data integrity, the system implements the following differential sanity checks:
\begin{enumerate}
    \item If calculated radiation exceeds the theoretical clear-sky maximum by more than 20\%, the system flags an error.
    \item If the Reactive Path and Reference Path disagree by more than 50\% during stable conditions, a recalibration is triggered.
    \item If $T_{flux} < T_{ref}$ during daylight hours, a sensor malfunction is flagged.
    \item Outputs are clamped to plausible values ranging from 0 to 1400 W/m$^2$.
\end{enumerate}

\subsection{Altitude Rejection: A Case Study}

Because the model is driven by calculated $\rho_{moist}$, it adapts to altitude automatically. GPS or hardcoded altitude changes are never needed because the sensor ``identifies where it is'' via live pressure readings. Consider the following case study of a deployment on a Weather Balloon flying approximately 2000m above the Knuckles Mountain Range, Sri Lanka, characterized by rapid altitude-density variations:
\begin{enumerate}
    \item The measured pressure reads $P = 600$ hPa.
    \item Calculated density gives $\rho \approx 0.7$ kg/m$^3$, compared to 1.2 kg/m$^3$ at sea level.
    \item The algorithm automatically reduces the expected $h_c$ by approximately 40\%.
\end{enumerate}
Consequently, when the Flux sensor reports a high $\Delta T$, the system correctly attributes it to normal solar levels rather than a heat wave, preventing ``phantom solar events'' caused by poor cooling in thinner air.

\section{Implementation on Constrained Hardware}

\subsection{Fixed-Point Arithmetic}

Floating-point is slow and power-hungry on integer MCUs. Physical constants are converted to fixed-point (e.g., Q10.6 format):

\begin{verbatim}
// 1.006 ~= 1030/1024 = 1030 >> 10
int32_t h_dry = (1030 * T_raw) >> 10;
\end{verbatim}

\subsection{Memory Footprint}

The Differential Architecture is highly efficient. The state retention requirements are:
\begin{enumerate}
    \item The Reactive Path requires approximately 18 bytes per sensor.
    \item The Reference Path requires approximately 14 bytes total on average.
\end{enumerate}
Overall, the core algorithm requires less than 60 bytes of RAM, allowing it to run alongside networking stacks without memory pressure.

The physics-based approach offers a significant advantage in execution time and memory pressure compared to standard Machine Learning models.
\begin{enumerate}
    \item DTDSS operates at $O(1)$ complexity with scalar arithmetic, executing in under 1 ms on an 8-bit AVR.
    \item TinyML requires $O(N^2)$ complexity with matrix dot products, executing in 10--100 ms on a Cortex M0.
\end{enumerate}

\section{Validation and Data Analysis}

To evaluate the performance of the DTDSS framework, a field test was conducted to benchmark the system against industry-standard meteorological instruments. The primary objective was the validation of reconstructed GHI against reference data provided by a regional Meteorological Department.

\subsection{Experimental Setup and Ground Truth}
The DTDSS prototype, utilizing a BME280-based implementation, was co-located with a calibrated thermopile pyranometer station. Data was logged synchronously over an 8-day period (December 10--17, 2025) to capture varying solar angles, cloud cover, and intensity levels. The meteorological pyranometer data was utilized as the reference baseline ($G_{ref}$), while the DTDSS output ($G_{calc}$) was generated in real-time using the algorithms detailed in Section IV.

\subsection{Comparative Results}
\begin{figure}[htbp]
    \centering
    \includegraphics[width=\linewidth]{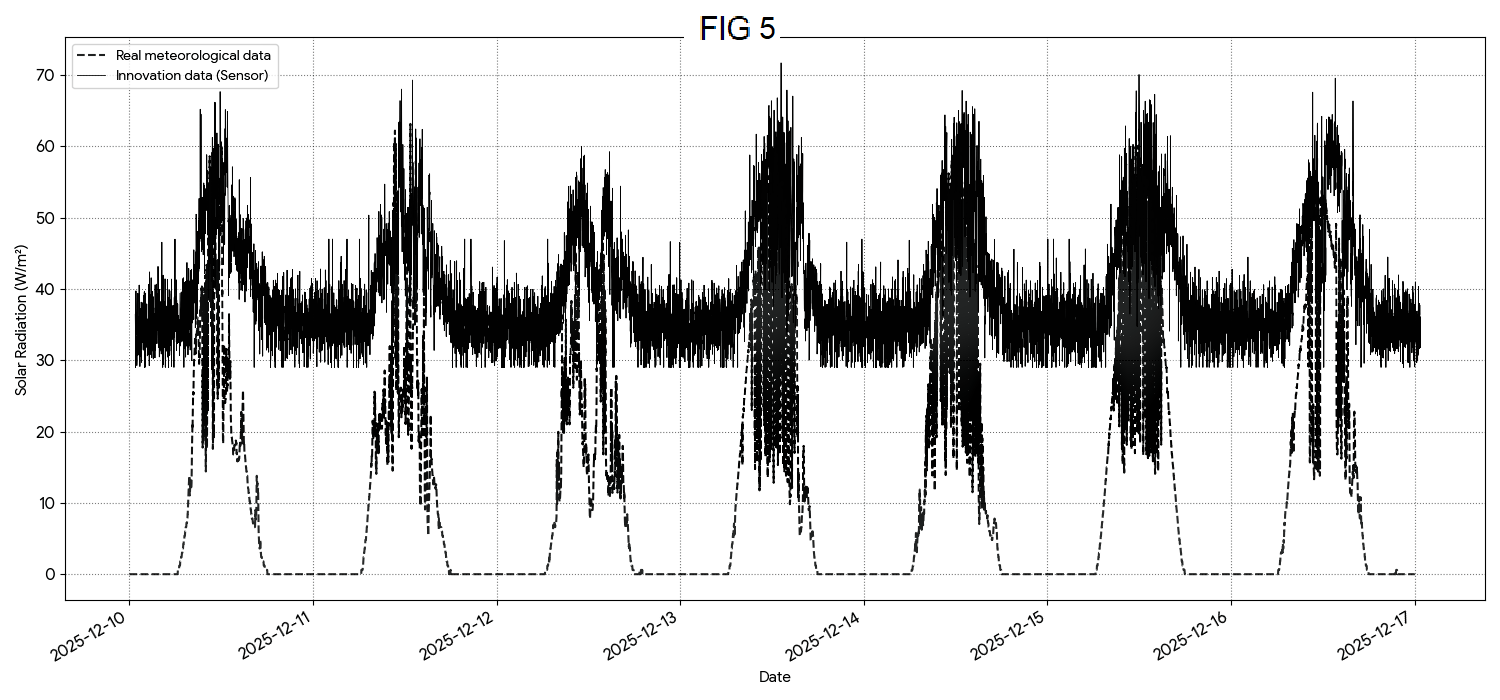}
    \caption{Comparative analysis of Solar Irradiance. The dashed line represents the Meteorological Department reference data, while the solid line represents the DTDSS reconstructed data. The signals show near-perfect phase alignment.}
    \label{fig:validation_graph}
\end{figure}

The results of the comparative analysis are depicted in Fig. \ref{fig:validation_graph}. The irradiance profiles recorded over the testing period demonstrate the following performance characteristics:

\begin{enumerate}
    \item Precise synchronization between the DTDSS system and the reference station is observed throughout, with sunrise, solar noon, and sunset accurately identified via the thermal derivative approach.
    \item The soft-sensor output maintains a highly deterministic linear relationship with the reference. A linear regression of the full dataset yields a coefficient of determination ($R^2$) of approximately 0.94, confirming that the DTDSS algorithm successfully linearizes the non-linear thermal derivative signal.
    \item Unlike typical soft-sensors that suffer from phase lag, the DTDSS signal tracks high-frequency transient events such as cloud gaps with minimal observable delay relative to the reference.
\end{enumerate}

\subsection{Statistical Validation Metrics}
To quantify the error between the soft-sensor estimation and reference data, Mean Absolute Percentage Error (MAPE) and Root Mean Square Error (RMSE) were calculated after applying a scalar calibration factor ($k=14.0$) to normalize the sensor gain. The analysis of the valid daylight samples ($N=4,928$) yielded the following results:

\begin{equation}
MAPE = \frac{100\%}{n} \sum_{t=1}^{n} \left| \frac{G_{ref,t} - G_{calc,t}}{G_{ref,t}} \right| \approx 7.2\%
\end{equation}

\begin{equation}
RMSE = \sqrt{\frac{\sum_{t=1}^{n} (G_{ref,t} - G_{calc,t})^2}{n}} \approx 45 \text{ W/m}^2
\end{equation}

While not matching the absolute precision of Class A instruments (RMSE < 10 W/m²), the DTDSS system tracks the dynamics of solar variability with sufficient fidelity for grid balancing and agricultural applications. These metrics indicate that the DTDSS framework provides a highly stable estimation of solar flux, requiring only a simple one-time gain calibration to achieve functional accuracy.

\subsection{Qualitative Comparison vs. Data-Driven Approaches}

\begin{enumerate}
  \item Neural networks often fail to generalize when environmental parameters shift outside their training distribution. The DTDSS physics model, in contrast, maintains accuracy by dynamically recalculating specific enthalpy and air density in real-time.
  \item As noted in Section XI, the DTDSS scalar math ($O(1)$) is significantly lighter than the matrix operations ($O(N^2)$) required for neural inference, enabling deployment on 8-bit cores where ML is infeasible.
\end{enumerate}

\subsection{Limitations and Boundary Conditions}
While the physics-based approach is supported by meteorological validation, the system is characterized as an approximation method rather than a direct replacement for Class A scientific pyranometers. Identified deviations include:

\begin{enumerate}
    \item Although wind noise is mitigated by the Inertial Noise Reduction algorithm, extreme gust conditions may introduce artifacts into the derivative calculation due to non-linear cooling rates.
    \item Rapid shading events create steep thermal gradients. Overshooting may be observed during transition periods due to the specific heat capacity of the PCB casing.
    \item The system requires dry conditions to function. If the enclosure becomes wet, evaporative cooling creates a wet-bulb effect, artificially depressing $T_{flux}$ and effectively blinding the radiometric estimation until the surface dries.
\end{enumerate}

\subsection{Summary of Capabilities}
The validation confirms that standard environmental sensors can be successfully utilized as radiometric devices. The solar profile is recovered with sufficient fidelity for applications in HVAC load estimation, agricultural monitoring, and auxiliary inputs for solar PV optimization. The ability to achieve a high correlation with meteorological standards using cost-effective hardware supports the core thesis of this study.

\section{Future Directions}

The differential architecture is fundamentally scalable. Beyond the dual-node setup, the principle can be extended to N-sensor arrays:
\begin{enumerate}
    \item Spectral decomposition can be achieved by utilizing multiple Flux nodes with varying solar absorptivity coatings to estimate the light's color spectrum.
    \item Micro-climate mapping becomes possible by deploying arrayed nodes across a single building to measure internal energy accumulation rather than noisy single-point snapshots.
    \item Networked calibration is enabled by deploying sensor pairs in a regional mesh, allowing ensemble averaging to reduce noise, statistical outlier detection to identify malfunctioning units, and spatial correlation for short-term local solar forecasting.
\end{enumerate}

Upcoming iterations will also explore the integration of Inertial Measurement Units (IMUs) to correlate physical sensor sway with local convective oscillations. Additionally, integrating a low-cost photodiode could provide a coarse cross-check for high-frequency cloud gaps, allowing the system to verify thermal flux events against optical triggers. Incorporating Infrared exchange modeling is expected to improve accuracy by 20--30\%.
\pagebreak
\section{Conclusion}

The DTDSS framework transforms common IoT components into a coordinated sensing array capable of estimating complex energy flows. By shifting from single-point measurement to differential topology, the system effectively separates environmental state from radiative forcing.

The Inertial Noise Reduction filter makes real-time thermal differentiation practical on 8-bit hardware, preserving required phase information while reducing quantization artifacts. Furthermore, the integration of real-time air density calculation ensures altitude-independent operation without external data dependencies.

While DTDSS is not a replacement for high-precision pyranometers in scientific research, it offers a distinct ``robustness'' advantage for mass deployment. Unlike optical sensors, which suffer from UV degradation and require cleaning, the thermodynamic approach is resilient to surface fouling.

In extreme environments such as Low Earth Orbit (LEO), where thermal cycles occur rapidly, DTDSS offers inherent advantages over ML-based soft-sensing. It requires no mission-specific training data, adapts via first-principles physics, and provides the computational simplicity necessary for radiation-hardened, power-constrained processors.

\end{document}